\documentclass[preprint,showpacs,preprintnumbers,amsmath,amssymb]{revtex4-1}
\usepackage{amsmath}
\usepackage{amssymb}
\usepackage{graphicx}
\usepackage{dcolumn} 
\usepackage{bm} 
\usepackage[hypertex]{}
\usepackage{longtable}
\usepackage[usenames]{color}
\usepackage[normalem]{ulem}

\begin{document}
\title{Surface and bulk Fermiology and band dispersion in non-centrosymmetric BiTeI}
\author{Gabriel Landolt$^{1,2}$}
\author{Sergey V. Eremeev$^{3,4}$}
\author{Yury M. Koroteev$^{3,4}$}
\author{Bartosz Slomski$^{1,2}$}
\author{Stefan Muff$^{1}$}
\author{Masaki Kobayashi$^{2}$}
\author{Vladimir N. Strocov$^{2}$}
\author{Thorsten Schmitt$^{2}$}
\author{Ziya S. Aliev$^{5}$}
\author{Mahammad B. Babanly$^{5}$}
\author{Imamaddin R. Amiraslanov$^{6}$}
\author{Evgueni V. Chulkov$^{7}$}
\author{J\"urg Osterwalder$^{1}$}
\author{J. Hugo Dil$^{1,2}$}
\affiliation{
$^{1}$Physik-Institut, Universit\"at Z\"urich, Winterthurerstrasse 190, 
CH-8057 Z\"urich, Switzerland 
\\ 
$^{2}$Swiss Light Source, Paul Scherrer Institut, CH-5232 Villigen, 
Switzerland
\\ 
$^{3}$Institute of Strength Physics and Materials Science, Siberian Branch, Russian Academy of Sciences, Akademicheskii pr. 2/1, Tomsk, 634021
Russia
\\ 
$^{4}$Tomsk State University, Tomsk, 634050 Russia
\\
$^{5}$Baku State University, General and Inorganic Chemistry Department, Baku, Azerbaijan
\\
$^{6}$Institute of Physics, Azerbaijan National Academy of Science, Baku, Azerbaijan
\\
$^{5}$Donostia International Physics Center (DIPS) and CFM, Centro Mixto CSIC-UPV/EHU, Departamento de F\'{i}sica de Materiales, UPV/EHU, 20080 San Sebasti\'{a}n, Spain
}
\date{\today}
\begin{abstract}
BiTeI has a layered and non-centrosymmetric structure where strong spin-orbit interaction leads to a giant spin splitting in the bulk bands. Here we present high-resolution angle-resolved photoemission (ARPES) data in the UV and soft x-ray regime that clearly disentangle the surface from the bulk electronic structure. Spin-resolved UV-ARPES measurements on opposite, non-equivalent surfaces show identical spin structures, thus clarifying the surface state character. Soft x-ray ARPES data clearly reveal the spindle-torus shape of the bulk Fermi surface, induced by the spin-orbit interaction.
\end{abstract}


\maketitle
The breaking of inversion symmetry and its influence on the spin structure of surface states under action of spin--orbit interaction (SOI) has been extensively studied in recent years \cite{Dil:2009review,Hirahara07}. The main finding is that the surface states become spin-split according to the Rashba model \cite{Bychkov:1984} resulting in two spin-polarized concentric Fermi contours. The lack of inversion symmetry in the bulk crystal structure is expected to induce a spin splitting with a more complex band- and spin-structure. Combined with strong SOI the Fermi surface can take the shape of a torus \cite{Cappelluti:2007}. For non-centrosymmetric superconductors such as for example CePt$_3$Si \cite{Bauer:2004} this peculiar band structure is expected to result in topologically protected spin polarized edge states reminiscent of Majorana modes \cite{Brydon:2011}.

Recently, an ARPES and spin-resolved ARPES study by Ishizaka \emph{et al.} \cite{Ishizaka:2011} proposed that the semiconductor BiTeI features a very large spin-splitting, arising from the broken inversion symmetry in the crystal bulk and a strong SOI. Theoretical work based on the perturbative $\mathbf{k}\cdot\mathbf{p}$ formalism linked the unusually large spin splitting in BiTeI to the negative crystal field splitting of the top valence bands \cite{Bahramy:2011}. Optical transition measurements \cite{Lee:2011} are in accordance with the giant bulk spin-splitting of the gap defining valence and conduction bands predicted by first principle calculations \cite{Ishizaka:2011,Bahramy:2011}. In addition it was shown in recent theoretical work that BiTeI can become a topological insulator under action of hydrostatic pressure \cite{Bahramy:2012}, and thus is closely related to non-centrosymmetric topological superconductors.

The present study provides first band mapping of a system without bulk inversion symmetry and giant SOI by the example of BiTeI, featuring a three-dimensional Rashba splitting of the bulk bands. Further it is shown that the Rashba-split state observed for this material in the UV photon energy regime is not a quantum well state \cite{Ishizaka:2011} but rather a surface state, using a simple symmetry argument based on spin-resolved ARPES (SARPES) measurements, which is confirmed by first principle calculations.

All measurements were performed at the Swiss Light Source of the Paul-Scherrer-Institut. The SARPES data was measured with the Mott polarimeter at the COPHEE endstation \cite{Hoesch:2002} of the Surface and Interface Spectroscopy beamline at a photon energy of 24 eV. The spin-integrated data at photon energies 20-63 eV were taken at the high-resolution ARPES endstation at the same beamline. The soft x-ray ARPES data were taken at the SX-ARPES endstation of the ADRESS beamline at photon energies of 310-850 eV. All spin-integrated measurements were performed at a sample temperature of 11 K and a base pressure lower than $10^{-10}$ mbar, the SARPES data was taken at 20 K.

For structural optimization and electronic band calculations we employed density functional theory (DFT) with the generalized gradient approximation (GGA) of Ref. \cite{PBE} for the exchange-correlation (XC) potential as implemented in VASP \cite{VASP}. The interaction between the ion cores and valence electrons was described by the projector augmented-wave method \cite{PAW}. The Hamiltonian contained scalar relativistic corrections, and the SOI was taken into account by the second variation method \cite{KH}.

\begin{figure}[t]
\begin{center}
\includegraphics[width=0.6\textwidth]{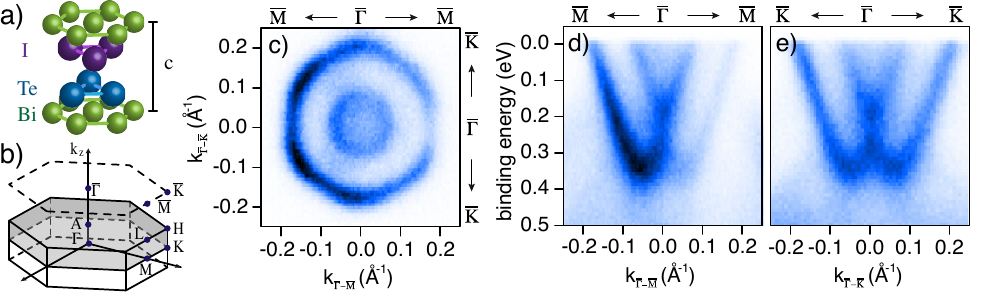}
\caption{(color online) (a) Crystal structure and (b) Brillouin zone of BiTeI. (c) Fermi surface map at 24 eV photon energy. (d) Band dispersion along $\overline{\Gamma}$--$\overline{\text{M}}$ and (e) $\overline\Gamma$--$\overline{\text{K}}$ obtained at the same photon energy.}
\label{fig:FSM}
\end{center}
\end{figure}

Figs. \ref{fig:FSM}(a,b) illustrate the hexagonal crystal structure and the Brillouin zone of BiTeI. The crystal is built up of alternating layers of bismuth, tellurium and iodine atoms stacked along the hexagonal axis. The continuous stacking order of the layers of the three atomic species breaks the inversion symmetry. DFT calculations have shown that the absence of inversion symmetry allows the strong SOI to lift the spin degeneracy of every band away from the time reversal invariant momenta ($\Gamma$,A,M,L). In particular, the bulk conduction band minimum (CBM) and the valence band maximum (VBM) are shifted away from the A-point at the Brillouin zone boundary towards the H- and L-points, resulting in a large Rashba-type spin splitting \cite{Ishizaka:2011}.

The ARPES measurement in Fig.~\ref{fig:FSM}(c) shows a section through the Fermi surface in the momentum plane parallel to the surface taken at a photon energy of 24eV. The two concentric rings reflect a typical Rashba-type band structure above the band crossing point. A slight deviation from the free-electron-like circular shape is observed for the outer ring, which is of hexagonal shape with the corners pointing towards the $\overline\Gamma$--$\overline{\text{K}}$ direction. The band maps in $\overline\Gamma$--$\overline{\text{M}}$ and $\overline\Gamma$--$\overline{\text{K}}$ direction in Fig.~\ref{fig:FSM}(d,e) show a nearly linear dispersion away from the parabola apices. The momentum splitting is slightly anisotropic but close to $2k_0\simeq 2\cdot0.057\text{\AA}^{-1}$.

Due to the broken inversion symmetry an untwinned BiTeI crystal has two different terminations. The weak bonding between the Te and I layers provides a natural cleaving plane, therefore the termination layer is either purely formed of iodine or tellurium atoms depending on the stacking order of the underlying layers. Despite the lacking inversion symmetry in the crystal structure the spin-integrated bulk band structure is inversion symmetric due to the time reversal symmetry. However, the spin structure of crystals with opposite stacking order is different as reflected in a reversal of the helicity of the spin-split bulk conduction band.

\begin{figure}[t]
\begin{center}
\includegraphics[width=0.8\textwidth]{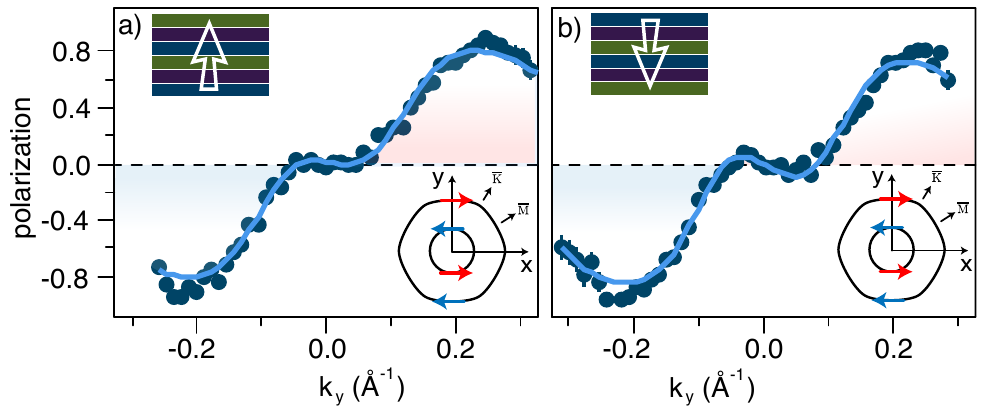}
\caption{(color online) (a) In-plane tangential spin polarization ($P_x$) at the Fermi level along  $\overline{\Gamma}$--$\overline{\text{M}}$ at 24 eV photon energy. (b) Same as (a) but sample is mounted with reversed c-axis. The lower right insets show the spin helicities at the Fermi surface.} 
\label{fig:SpinCompare}
\end{center}
\end{figure}

Figures~\ref{fig:SpinCompare}(a) and (b) compare the measured in-plane spin polarization at a photon energy of 24eV along $\overline{\Gamma}$--$\overline{\text{M}}$ for two samples taken from the same batch but mounted with inverted c-axis. 
The spin polarization data reflect the structure of a large Rashba-type spin splitting with a high in-plane tangential polarization and clockwise helicity of the outer bands. The inner bands are strongly overlapping with the adjacent bands resulting in a small polarization amplitude around normal emission (see Ref. \cite{Meier:2008,SOM} for details on the fitting procedure). Despite the inverted crystal c-axis, the measured spin helicities do not reverse and no significant changes in the momentum dependence of the polarization can be observed.

\begin{figure*}[htb]
\begin{center}
\includegraphics[width=\textwidth]{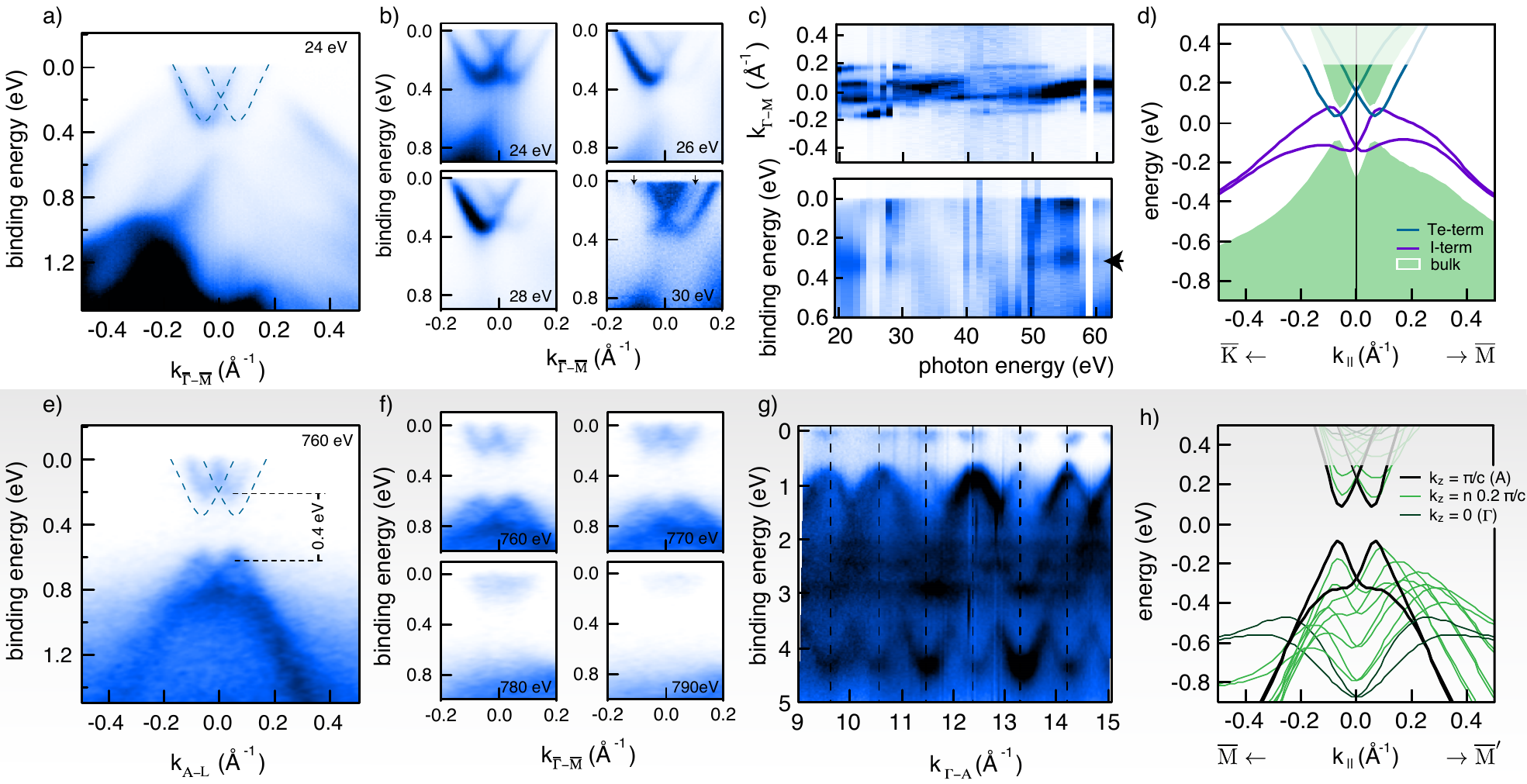}
\caption{(color online) (a)-(b) High resolution ARPES band map of the surface state at 24-30eV photon energy. (c) Fermi surface map along $\overline\Gamma$--$\overline{\text{M}}$ as a function of photon energy and map of the parabola minimum at $0.06\text{ \AA}^{-1}$. (d) Slab calculations for iodine and tellurium termination. For both terminations a surface state appears in the bulk band gap. (e) SX-ARPES band map of the bulk state at 760eV photon energy ($16^\text{th}$ zone boundary). (f) Band maps at 760eV--790eV photon energy. (g) Band map along $\Gamma$--A (i.e. $k_\parallel=0$), the dashed lines indicate the Brillouin zone boundaries. (h) DFT bulk spectra for a set of $k_z$-values along $\overline{\text{M}}$--$\overline\Gamma$--$\overline{\text{M}}'$.}
\label{fig:kzCompare}
\end{center}
\end{figure*}

The dispersive behavior in the out-of-plane momentum ($k_z$) direction of these conduction bands is examined by photon energy dependent band maps as shown in Fig.~\ref{fig:kzCompare}(a,b). Aside from strong intensity variations due to matrix element effects the bands do not change as a function of photon energy. Taking into account that at these low photon energies half a Brillouin zone is swept by changing the photon energy by roughly 10 eV, the states can be said to not disperse in $k_z$ at all. There is neither a significant photon energy dependence in the Fermi momenta nor in the parabola apex around 0.33 eV, as can be seen in the upper and lower panel in Fig.~\ref{fig:kzCompare}(c).
The absence of $k_z$-dispersion is a typical feature of a two-dimensional state such as a quantum well state or a surface state. However, the stacking-independent spin helicity fundamentally contradicts the expected properties of a bulk-derived quantum well state. We conclude that the observed state is actually a surface state that overshadows the bulk states. At 30 eV photon energy the bulk states are weakly visible at lower binding energies and momenta (the Fermi momenta are indicated by arrows in the lower right panel in Fig.~\ref{fig:kzCompare}(b)).

Measurements in the soft x-ray photon energy range corroborate this interpretation. At these energies the spectroscopic intensities shift from the surface states to the bulk states due to the highly localized nature of the surface state and the larger probing depth. Fig.~\ref{fig:kzCompare}(e) shows a band map along A--L measured at 760eV photon energy. In contrast to the surface state, the conduction band minimum is about 100meV lower in binding energy compared to the low photon energy measurement in Fig.~\ref{fig:kzCompare}(a), furthermore the momentum splitting is about 20\% smaller. In accordance with previous optical measurements the bulk band gap is 400 meV \cite{Lee:2011}. That these bands observed in the soft x-ray photon energy range are indeed bulk bands is evident from the three dimensional behavior of the dispersion in Figs.~\ref{fig:kzCompare}(f). Both for the valence band and the conduction band a clear $k_z$ dispersion is observed. At 760eV photon energy, corresponding to the $16^\text{th}$ A-point in the Brillouin zone, the CBM binding energy is largest and decreases away from the A-point and eventually disperses above the Fermi level around 800eV photon energy. Fig.~\ref{fig:kzCompare}(g) shows the dispersion along the $\Gamma$--A direction, i.e. for $k_\parallel=0$, measured by sweeping photon energies from 310eV to 850eV. Because of the non-negligible photon momentum transferred to the photoelectrons at the used photon energies, the measured electron momenta have been corrected for the photon momentum projected on the particular electron momentum axis. The $k_z$-dispersion of the conduction band follows the periodicity of the valence bands, so the highest CBM binding energies occur periodically at all A-points ($k_z = (2n+1)\pi/c$), indicated by the vertical dashed lines. In the entire measured energy range no photoemission intensity can be observed from the surface state due to the lower photoemission cross-section.

First principle calculations nicely reproduce the observations. The calculated $k_z$-de\-pen\-dence of the lowest conduction band and highest valence band are shown in Fig.~\ref{fig:kzCompare}(h) along $\overline{\text{M}}$--$\overline{\Gamma}$--$\overline{\text{M}}$'. At the A-point the calculation perfectly matches the measured spectrum at 760eV in Fig.~\ref{fig:kzCompare}(d), apart from a smaller calculated band gap, which is a typical problem of DFT calculations.

The origin of the two surface states visible in UV ARPES can be understood from DFT calculations for both Te- and I-terminated surfaces. To calculate the spectrum of the Te-terminated surface we used a 24 layer slab with a free tellurium surface on one slab side and a hydrogen-passivated iodine termination on the other side. The tellurium surface hosts the Rashba-split electron-like surface state within the bulk band gap. This state lies 90 meV below the conduction band bottom and has a 20\% larger momentum splitting than the bulk conduction band, which is in excellent agreement with the observed photoemission data. On the other hand a hole-like surface state appears at the I-terminated surface (in this case the Te-terminated side of the slab is passivated). In Fig.~\ref{fig:kzCompare}(d) we show both spin-split surface states at the Te- and I-terminated surfaces. In photoemission data in Fig.~\ref{fig:kzCompare}(a) a hole-like state seems indeed to cross the gap and overlap with the electron-like surface state in agreement with the theoretical predictions. The measured spin-helicity of the Te-derived surface state is reversed compared to the Au(111) surface state \cite{Hoesch:2004}. This is confirmed by layer-resolved spin calculations \cite{SOM}, where similar to the Dirac state of Bi$_2$Te$_3$ and PbBi$_4$Te$_7$ a reversed spin contribution of the outermost Te-layer with respect to next atomic layers is observed \cite{Eremeev:2012}.

In all measurements on various samples and for both crystal mounting orientations the surface states of both terminations are present as in Fig.~\ref{fig:kzCompare}(a) and all samples show the same spin structure. This leads to the conclusion that both surface terminations are present in domains of sizes smaller than the synchrotron light spot. This is only possible if both stacking orders are equally present in the sample, due to a large number of stacking faults. This explains why no bulk contribution to the spin polarization could be observed, since in this case any spin polarization of one domain is expected to be cancelled by an other with opposite stacking order.

\begin{figure}[t]
\begin{center}
\includegraphics[width=0.8\textwidth]{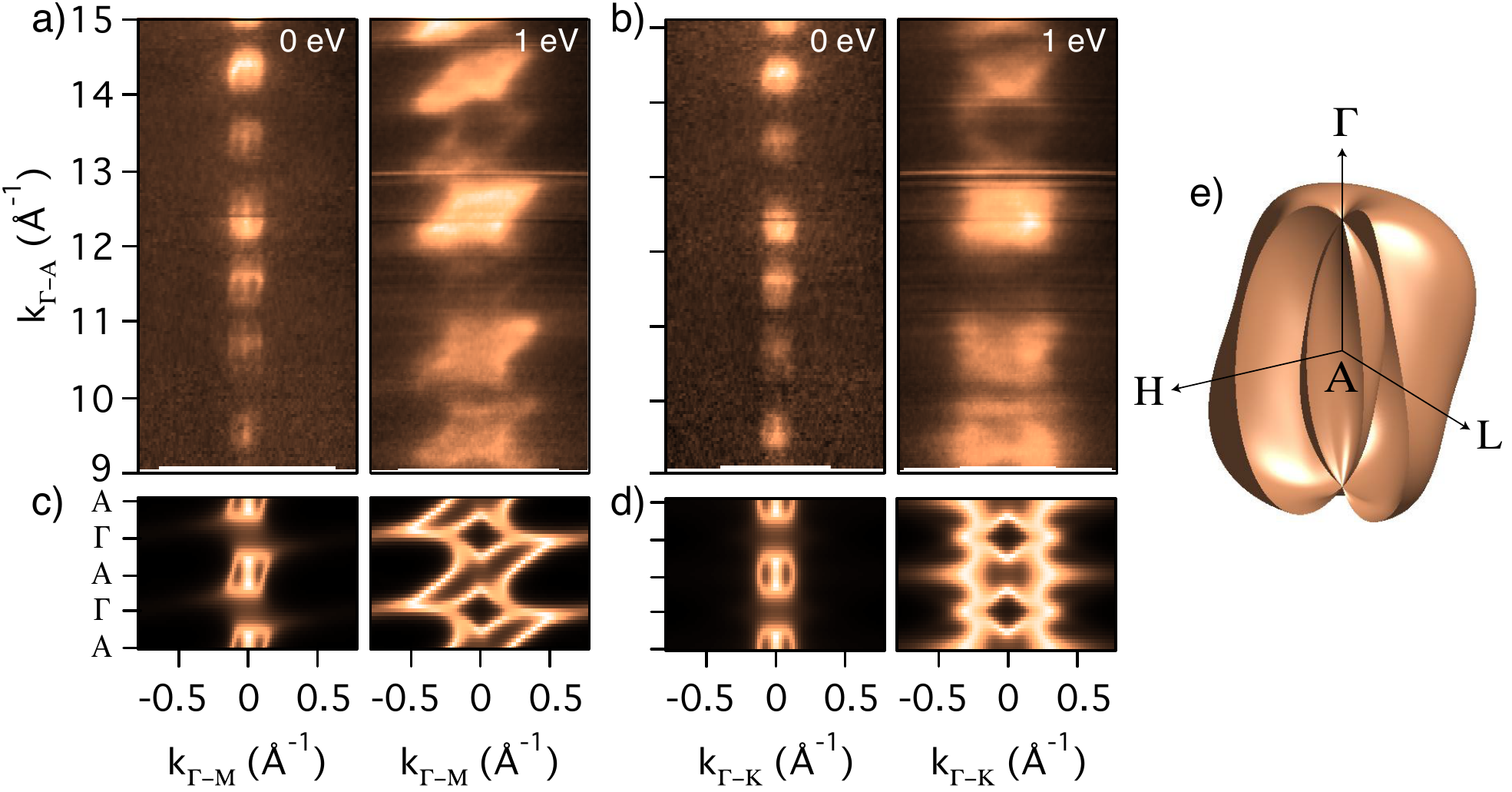}
\caption{(color online) (a,b) SX-ARPES photon-energy dependent constant binding-energy maps of the lowest bulk conduction band at the Fermi level and of the highest valence band at 1 eV binding energy in the (a) $\Gamma$--M--A--L and (b) $\Gamma$--K--A--H plane. (c,d) DFT calculations corresponding to the directions and energies in (a) and (b). (e) Schematic representation of the 3D Fermi surface.}
\label{fig:kzSymmetry}
\end{center}
\end{figure}

Having identified the surface state contributions to the ARPES spectra, we will now further discuss the properties of the bulk band structure. In the in-plane direction the Fermi surface around the A-point is described by two concentric rings, resembling the Fermi surface of the electron-like surface state, while it disappears around the $\Gamma$-point. In the out-of-plane direction the Fermi surface takes the form of two loops, intersecting at the $\Gamma$-A line (Fig.~\ref{fig:kzSymmetry}(d)). The left panels of Figs.~\ref{fig:kzSymmetry}(a) and (b) show the Fermi surface in the $\Gamma$--M--L--A and $\Gamma$--K--H--A plane, respectively. The right panels display constant energy cuts in the same momentum planes at a binding energy of 1eV cutting through the valence band close the lower gap edge.

The measured constant energy cuts are well reproduced by DFT calculations. In Fig.~\ref{fig:kzSymmetry}(c) sections through the calculated Fermi surface in the momentum planes corresponding to the measurements in Figs. \ref{fig:kzSymmetry}(a,b), about 275meV above the CBM, are shown in the top panels, likewise constant energy cuts through the valence band 485meV below the VBM in the lower panels. The lifetime band broadening has been manually added. These data complete the picture of the complex 3D Rashba-type Fermi surface taking the form of a spindle torus (Fig.~\ref{fig:kzSymmetry}(d)) distorted according to the crystal symmetry. In a topologically non-trivial system with a similar band structure a zero-energy mode will connect the crossing points in the $\Gamma$--A direction when a superconducting gap is opened \cite{Brydon:2011}.

To conclude, we have clarified the surface state character of the two-dimensional Rashba-split conduction-band state at the Te-terminated surface of BiTeI. Further, we have proven by soft x-ray ARPES measurements that also the bulk conduction band shows a three-dimensional Rashba-type band splitting. Our measurements provide a direct mapping of the Fermi surface of a 3D non-centrosymmetric crystal structure. The large spin splitting in both the bulk and surface states makes this material promising for spin transport applications.

We thank E. Razzoli, M. Radovic, F. Dubi, M. Kropf, and C. Hess for support during the measurements. This work was supported by the Swiss National Science Foundation.

\bibliographystyle{apsrev4-1}

\end{document}